\documentclass{article}

\usepackage{graphicx}
\usepackage{amssymb,amsfonts,amsmath,amsthm}
\usepackage[round]{natbib}
\usepackage{setspace}

\newcommand{\dkl}{\delta_{\mathrm{KL}}}
\newcommand\ring[1]{\mathaccent23{#1}}
\newcommand{\ps}{\Delta}
\newcommand{\psint}{\ring{\Delta}}
\newcommand{\paths}{\mathcal{P}}
\newcommand{\pmin}{p_{\min}}

\newcommand{\SBsection}[1]{\vspace{.2cm} \noindent \begin{center}
    \textsc{#1} \end{center} \vspace{-.05cm}}
\newcommand{\SBsubsection}[1]{\vspace{.0cm} \noindent \begin{center}
    \textit{#1} \end{center} \vspace{-.1cm}}
\newcommand{\SBsubsubsection}[1]{\vspace{.4cm} \noindent \begin{center}
  \textsc{\small #1} \end{center} \vspace{0cm}}

\newenvironment{sysbiol}{\begin{raggedright} \parindent=0.5in}{\end{raggedright}}

\newtheorem{thm}{Theorem}

\newtheorem{lem}[thm]{Lemma}

\newtheorem{prop}[thm]{Proposition}

\theoremstyle{definition}

\theoremstyle{definition}

\theoremstyle{definition}

\begin{document}

\begin{sysbiol}

\begin{center}

\emph{Running head:} 

PHYLOGENETIC MIXTURES ON A SINGLE TREE CAN MIMIC ANOTHER TOPOLOGY

\vspace{.5cm}

\LARGE{Phylogenetic mixtures on a single tree can mimic a tree of another topology}

\vspace{1cm}
\large{
Frederick A. Matsen and Mike Steel \\
\textsl{
Biomathematics Research Centre\\ 
University of Canterbury\\ 
Private Bag 4800\\ 
Christchurch, New Zealand
}}

\vspace{1cm}
\textsl{
Corresponding Author:\\
Frederick A. Matsen\\
phone: +64 3 364 2987 x7431\\
fax: +64 3 364 2587\\
email: ematsen@gmail.com\\
}

\end{center}

\vspace{1cm} 

{\noindent Keywords: Phylogenetics; Mixture Model; Sequence Evolution;
Model Identifiability}

\newpage

\begin{spacing}{1}
 
\SBsection{Abstract}
Phylogenetic mixtures model the inhomogeneous molecular evolution
commonly observed in data.  The performance of phylogenetic
reconstruction methods where the underlying data is generated by a
mixture model has stimulated considerable recent debate.  Much of the
controversy stems from simulations of mixture model data on a given
tree topology for which reconstruction algorithms output a tree of a
different topology; these findings were held up to show the
shortcomings of particular tree reconstruction methods.  In so doing,
the underlying assumption was that mixture model data on one topology
can be distinguished from data evolved on an unmixed tree of another
topology given enough data and the ``correct'' method.  Here we show
that this assumption can be false.  For biologists our results imply
that, for example, the combined data from two genes whose phylogenetic
trees differ only in terms of branch lengths can perfectly fit a tree
of a different topology.

\newpage

It is now well known that molecular evolution is heterogeneous, i.e.
that it varies across time and position \citep{pmid8752001}.
A classic example is stems and loops of
ribosomal RNA: the evolution of one side of a stem is strongly
constrained to match the complementary side, whereas for loops
different constraints exist \citep{pmid8798341}. Heterogeneous evolution between genes is
also widespread, where even the general features of evolutionary
history for neighboring genes may differ wildly \citep{pmid10830951}.
Presently it is not uncommon to use concatenated sequence data from
many genes for phylogenetic
inference \citep{pmid15084674}, which can lead to very high levels
of apparent heterogeneity \citep{pmid11062127}.
Furthermore, empirical evidence using the covarion model shows that
sometimes more subtle partitions of the data can exist, for which separate analysis
is difficult \citep{covarion-test}.

This heterogeneity is typically formulated as a mixture
model \citep{pmid15371247}. Mathematically, a phylogenetic mixture model is
simply a weighted average of site pattern frequencies derived from a
number of phylogenetic trees, which may be of the same or different
topologies. Even though many phylogenetics programs
accept aligned sequences as input, the only data actually used in the vast
majority of phylogenetic algorithms is the derived site pattern frequencies.
Thus, in these algorithms, any record of position is lost and
heterogeneous evolution appears identical to homogeneous
evolution under an appropriate phylogenetic mixture model.
For simplicity, we call a mixture of site pattern frequencies from two
trees (which may be of the same or different topology) a
\emph{mixture of two trees}; when the two trees have the same
underlying topology, the mixture will be called a \emph{mixture of
branch length sets on a tree}. 

Mixture models have proven difficult for phylogenetic
reconstruction methods, which have historically sought to find a
single process explaining the data. For example, it has been shown
that mixtures of two different tree topologies can mislead MCMC-based
tree reconstruction \citep{pmid16195459}. It is also known
that there exist mixtures of branch length sets on one tree which
are indistinguishable from mixtures of branch length sets on a tree of
a different
topology \citep{pmid8790461,stefankovic-vigoda,stefankovic-vigoda-SB}.
Recently, simulations of mixture models from ``heterotachous''
(changing rates
through time) evolution have been shown to cause reconstruction
methods to fail \citep{artifactual}.

The motivation for our work is the observation that both theory and
simulations have shown that in certain parameter regimes, phylogenetic
reconstruction methods return a tree topology different
from the one used to generate the mixture
data.
The parameter regime in this class of examples is similar to
that shown in Figure~1, with two neighboring pendant edges which
alternate being long and short. After mixing and reconstruction, these
edges may no longer be adjacent on the reconstructed tree. We call this 
\emph{mixed branch repulsion}.
This phenomenon has been
observed extensively in simulation \citep{pmid15496922,pmid15746012,pmid16209710,pmid16014870} and
it has been proved that certain
distance and maximum likelihood methods are susceptible to this
effect \citep{pmid8664540,stefankovic-vigoda,stefankovic-vigoda-SB}.
 Up to this point such results have been interpreted as pathological
behavior of the reconstruction algorithms, which
has led to a heated debate about which reconstruction
methods perform best in this situation \citep{pmid15922824,
  pmid15922825}.
Implicit in this debate is the assumption that a mixture of trees on
one topology gives different site pattern frequencies than that of an
unmixed tree of a different topology. This leads to the natural question of
how similar these two site pattern frequencies can be.

Here we demonstrate that mixtures of two sets of branch lengths on a tree of
one topology can exactly mimic the site pattern frequencies of a
tree of a different topology under the two-state symmetric model. In
fact, there is a precisely
characterizable (codimension two) region of parameter space where such
mixtures exist. Consider two quartet trees of topology $12|34$, as
shown in Figure~1. Label the pendant branches 1 through 4 
according to the taxon labels, and label the internal edge with 5. The first branch length
set will be written $t_1,\ldots,t_5$ and the second $s_1,\ldots,s_5$.
Now, if $k_1,\ldots, k_4$ satisfy the following system of
inequalities
\begin{eqnarray*}
  & k_1 > k_3 > k_4 > 1 > k_2, & \\
  & \frac{1-k_1^2}{k_1} \frac{1-k_4^2}{k_4} + \frac{1-k_2^2}{k_2}
  \frac{1-k_3^2}{k_3} > 0, & \\
  & \frac{k_1+k_4}{1+k_1 k_4} \cdot \frac{k_2+k_3}{1+k_2 k_3} > 1 & \\
\end{eqnarray*}
then they specify a class of examples of mixed branch repulsion.
More precisely, then there exist nonzero internal branch lengths $t_5$ and $s_5$,
mixing weights, and positive numbers $\ell_1, \ldots, \ell_4$ such
that if for $i=1,\dots,4$, $k_i = \exp\left(-2(t_i - s_i)\right)$ and
$t_i \geq \ell_i$, the corresponding mixture of two $12|34$ trees will
have the same site pattern frequencies as a single tree of the $13|24$
topology. We have
illustrated two examples of branch length sets satisfying these
criteria in Figure~1 and provided the corresponding branch lengths in
Table~1. 

The exact zone for mixed branch repulsion is described above and detailed in 
Proposition~\ref{prop:main}; here we present some simple necessary criteria
for mixed branch repulsion to occur.
First, note that except for the internal edge and a (typically small) lower
bound on pendant branch lengths, the relevant parameters are differences
of branch lengths between sets rather than absolute branch lengths
themselves. Given two branch length sets
with edges numbered as above, let
$d_i$ denote the difference between the branch lengths for edge $i$,
i.e. $t_i - s_i$.
Then (perhaps after changing the arbitrary numbering of the taxa) either 
$d_1 > d_3 > d_4 > 0 > d_2$ or 
$d_1 > 0 > d_3 > d_4 > d_2$ 
must be satisfied in order for mixed branch repulsion to occur. Thus,
for example, in one set of branch lengths
the pendant edge for taxa $1$ should be long and the pendant edge for
taxa $2$ should be short, while in the other set of branch lengths
these roles should be reversed. On the other hand, the branch lengths
for taxa $3$ and $4$ should be both long for one set and both short
for the other.
Additionally, at least one of the
two internal branch lengths needs to be relatively short. There are
other more complex criteria, but the above is necessary for exact mixed
branch repulsion to occur. However, as noted below, exact mixed branch
repulsion is not necessary to ``fool'' model based methods.

We believe that this similarity between site pattern frequencies
generated by mixtures of branch lengths on one tree and corresponding
unmixed frequencies on a different tree is what is leading to the mixed
branch repulsion observed in theory and
simulation. 
Furthermore, it is possible that even the simple
case  presented here is directly relevant to reconstructions from
data. First, it
is not uncommon to simplify the genetic code from the four standard
bases to two (pyrimidines versus purines) in order to reduce the
effect of compositional bias when working with genome-scale data on
deep phylogenetic relationships \citep{pmid15084674}. Second, when
working on such relationships concatenation of genes is
common \citep{pmid11062127}, for
which a phylogenetic mixture is the expected result. Finally,
the region of parameter space bringing about mixed
branch repulsion may become more extensive as the number of
concatenated genes increases.  Therefore in concatenated gene analysis
it may be worthwhile considering incongruence in terms of branch
lengths and not just in terms of
topology \citep{pmid14574403,pmid16490279}, as highly incongruent branch
lengths may produce artifactual results upon concatenation. Other
methods may be useful in this setting, such as gene order data, gene
presence/absence, or coalescent-based methods to infer the most likely
species tree from a collection of gene trees.

Mixed branch repulsion may be more difficult to detect than the usual
model mis-specification issues; in the cases presented here the mis-specified
single tree model fits the data perfectly. In contrast, although using the wrong mutation
model for reconstruction using maximum likelihood can lead to
incorrect tree topologies \citep{pmid15930156}, the
resulting model mis-specification can be seen from a poor
likelihood score. In the mixtures presented here, there is no
way of telling when one is in the mixed regime on one topology or an
unmixed regime on another topology.  Furthermore, any model selection
technique (including likelihood ratio tests, the Akaike Information
Criterion and the Bayesian Information Criterion) which chooses a
simple model given equal likelihood scores would, in this case, choose
a simple unmixed model. Thereby it would select a tree that is different
from the historically correct tree if the true process was generated
by a mixture model.

The derivation of the zone resulting in mixed branch repulsion
is a conceptually simple application of two of the pillars of theoretical
phylogenetics: the Hadamard transform
and phylogenetic invariants \citep{hendypenny89,semple-steel,felsenstein}. The
Hadamard transform is a closed form invertible transformation
(expressed in terms of the discrete Fourier transform) for
gaining the expected site pattern frequencies from the branch lengths and
topology of a tree or vice versa. Phylogenetic invariants
characterize
when a set of site pattern frequencies could be the expected site
pattern frequencies for a tree of a given topology. They are identities
in terms of the discrete Fourier transform of the site pattern
frequencies.  Therefore, to derive the above equations, we simply
insert the Hadamard formulae for the Fourier transform of pattern probabilities
into the phylogenetic invariants, then check to make sure the
resulting branch lengths are positive.

Similar considerations lead to an understanding of when it is
possible to mix two branch length sets on a tree to reproduce the site
pattern frequencies of a tree of the same topology
(Proposition~\ref{prop:same_mix} of Appendix). For a quartet, two
cases are possible. First, a pair of neighboring pendant branch lengths can be
equal between the two branch length sets of the mixture.
Alternatively, the sum of one pair of neighboring pendant branch
lengths and the difference of the other pair can be equal. For trees larger than
quartets, the allowable mixtures are determined by these restrictions
on the quartets (results to appear elsewhere).  For pairs of branch
lengths satisfying these criteria, any choice of mixing weights will
produce site pattern frequencies satisfying the phylogenetic
invariants.

Intuitively, one might expect that when two sets of branch lengths mix
to mimic a tree of the same topology, some sort of averaging property
would hold for the branch lengths. This is true for pairwise distances
in the tree but need not be the case for individual branches, as
demonstrated by Figure~2. In fact, it is possible to mix two sets of
branch lengths on a tree to mimic a tree of the same topology such
that a resulting pendant branch length is arbitrarily small while the
corresponding branch length in either of the branch length sets being
mixed stays above some arbitrarily large fixed value. 

The results in this paper shed some light on the geometry of
phylogenetic mixtures \citep{pmid11020305}.  As is
well known, the set of phylogenetic trees of a given topology forms a
compact subvariety of the space of site pattern
frequencies \citep{pmid15767777}. The first part of our work
demonstrates that there are pairs of points in one such subvariety
such that a line between those two points intersects a distinct
subvariety (see Figure~3). Therefore the convex hull of one subvariety has a region
of intersection with distinct subvarieties. This is stronger 
than the recently derived result by \citet{stefankovic-vigoda,stefankovic-vigoda-SB} that the convex hulls
of the varieties intersect.  The second part of our work shows that there exist
pairs of points in a subvariety such that the line between those
points intersects the subvariety.  Furthermore, it demonstrates that
when such a line between two points intersects the subvariety in a third
point, then a subinterval of the line is contained in the subvariety.

This geometric perspective can aid in understanding practical problems
of phylogenetic estimation. The question of
when maximum likelihood selects the ``wrong'' topology given mixture
data was initiated
by \citet{pmid8664540} who found a one-parameter space of such
examples under the two-state symmetric (CFN) model. Recently \citet{stefankovic-vigoda}
found a two-parameter space of such examples for the CFN model, and a
one-dimensional space of examples for the Jukes-Cantor DNA (JC) and
Kimura two and three parameter (K2P, K3P) models. A potential
criticism of these results is that because the set of examples has
lower dimension than the ambient parameter space one is unlikely to
encounter them in practice.

However, a simple geometric argument can show that the dimension of
the set of all such pathological examples is equal to the dimension of
the parameter space for all four of these models.
To see why this holds we first
recall the definition of the Kullback-Leibler divergence of
probability distribution $q$ from a second distribution $p$:
\[
\dkl(p,q) = \sum_i p_i \log \frac{p_i}{q_i}. 
\]
The $p$ vector is typically thought of as a data vector and the $q$
vector is typically the model data. Maximum likelihood seeks to find the 
model data vector $q$ which minimizes $\dkl(p,q)$.
Let $V_{12|34}$ be the set of all data vectors
which correspond exactly to trees of topology $12|34$, and similarly
for $V_{13|24}$.
For $V = V_{12|34}$ or $V_{13|24}$ let $\dkl(p,V)$ denote the divergence of $p$ from the
``closest'' point in $V$, i.e. the minimum of $\dkl(p,v)$ where $v$
ranges over $V$ . We show in
Lemma~\ref{lem:kl_cont} that this
function exists and is continuous across the set of probability
vectors $p$ with all components positive. 

Now, pick any of the above group-based models, and let $y$ be a
corresponding pathological mixture on $12|34$ for that model supplied by
Theorem~2 of \citet{stefankovic-vigoda}. 
Maximum likelihood chooses topology $13|24$ over $12|34$ for a
data vector $p$ exactly when
$\dkl(p,V_{13|24})$ is less than $\dkl(p,V_{12|34})$, therefore 
$\dkl(y,V_{13|24}) < \dkl(y, V_{12|34})$. By the properties of
continuous functions, this inequality also holds for all probability
vectors $y'$ close to $y$ which also have all components positive.
Therefore ML will choose $13|24$ over $12|34$ for all such $y'$.
Because the transformation taking branch length and mixing weight
parameters to expected site pattern frequencies is continuous, one can
change branch lengths and mixing weight arbitrarily by a small amount
and still have ML choose $13|24$ for the resulting data. This gives
the required full-dimensional space of examples.

We now indicate how our results fit into previous work on
identifiability and discuss prospects for generalization. 
For four-state models with extra symmetries such as the
Jukes-Cantor DNA model and the Kimura two-parameter model it is known
that there exist
linear phylogenetic invariants which imply identifiability of the
topology for mixture model data \citep{stefankovic-vigoda}.  The
topology is also identifiable for phylogenetic mixtures in which each
underlying process is described by an infinite state model
\citep{moste1, moste2} -- such processes may be relevant to data
involving rare (homoplasy-free) genomic changes.  Therefore the
pathologies observed here could not occur for those models.
Furthermore, \citet{allman-rhodes} have shown generic identifiability
(i.e.  identifiability for ``almost all'' parameter regimes) when the
number of states exceeds the number of mixture classes. As stated
above, the dimension of the set of examples presented here is of
dimension two less than the ambient space (even though the conditions
of the Allman and Rhodes work is not satisfied).  However, we note
that even when tree topology is generically identifiable (but not
globally identifiable) for some model, arguments similar to the above
can show that there exist positive-volume regions where the data is
closer to that from a tree of a different topology than a tree of the
same topology. 

A related though distinct question concerns identifiability under
mixture models when the data partitions are known.
For example, we may have a number of independent
sequence data sets for the same set of taxa, perhaps corresponding
to different genes. In this setting it may be reasonable to assume
that the sequence sites \emph{within} each data set evolve under the same
branch lengths (perhaps subject to some i.i.d. rates-across-sites
distribution), but that the branch lengths \emph{between} the data sets
may vary. The underlying tree topology may be the same or different
across the data sets, however let us first consider the case where
there is a common underlying topology. In the case where each data set
consists of sequences of length one we are back in the setting of
phylogenetic mixtures considered above.  However, for longer blocks of
sequences, we might hope to exploit the knowledge that the sequences
within each block have evolved under a common mechanism.  If the
sequence length within any one data set becomes large we will be able
to infer the underlying tree for that data set correctly, so the
interesting question is what happens when the data sets  provide only
`mild' support for their particular reconstructed tree.  
Assume that all (or nearly all) of the data sets contain sufficiently
many sites so that the tree reconstruction method $M$ positively favors the true
tree over any particular alternative tree.  By this we mean that $M$
returns the true tree with a probability that is greater by a factor of
at least $1+ \epsilon$ (with $\epsilon > 0$) than the probability that
$M$ returns each particular different tree. 
Then it is
easily shown that a majority rule selection procedure applied to the
reconstructed trees across the $k$ independent data sets will
correctly return the true underlying tree topology with a probability that goes
to $1$ as $k$ grows. Note that this claim holds generally, not just for the
two-state symmetric model.  Of course it is also possible that the
underlying tree may differ across data sets-- in the case of genes
perhaps due to lineage sorting \citep{deneganRosenberg06}-- in which
case the reconstruction question becomes more complex.
 
In a forthcoming article (Matsen, Mossel, and Steel 2007) we further
investigate identifiability of mixture models. Using geometric methods
we make some progress towards understanding how ``common''
non-identifiable mixtures should be for the symmetric and
non-symmetric two-state models; for mixtures of many trees they appear
to be quite common. A new combinatorial theorem 
implies identifiability for certain types of mixture models when
branch lengths are clock-like.  A simple argument shows
identifiability for rates-across-sites models. We also investigate
mixed branch repulsion for larger trees.

Many interesting questions remain. First of all, is exact mixed branch
repulsion an issue for any nontrivial model on four states?  
Also, what is the zone of parameter space for which a mixture of branch
lengths on a tree is closer (in some meaningful way) to the expected
site pattern frequencies of a tree of different topology than to those
for a tree of the original topology? How often does mixed branch
repulsion present itself given ``random'' branch lengths? Considering
the rapid pace of development in this field we do not expect these
questions to be open for long.

\nocite{mixedup07}

\newpage
\SBsubsubsection{Acknowledgments}

\begin{footnotesize}
The authors would like to thank 
Cecile An\'e, Andrew Roger, Jack Sullivan, and an anonymous
reviewer for comments which greatly improved the
paper. 
Dennis Wong provided advice on the figures, and
David Bryant's Maple code was used to check results.
Funding for this work was provided by the Allan Wilson Centre for
Molecular Ecology and Evolution, New Zealand.
\end{footnotesize}

\end{spacing}


\newpage
\bibliographystyle{sysbio}
\bibliography{mixtures}


\begin{spacing}{1}

\newpage
\SBsection{Appendix}

In this section we provide more precise statements and
proofs of the propositions in the text. The proofs will be presented in
the reverse order than they were stated in the main text--- first the fact
that it is possible to mix two branch lengths on a tree to mimic a tree of
the same topology, then that it is possible to mix branch lengths to mimic a
tree of a distinct topology.

As stated in the main text, the general strategy of the proofs is
simple: use the Hadamard transform to calculate Fourier transforms of
site pattern probabilities and then insert these formulas into the
phylogenetic invariants. These steps would become very messy except
for a number of simplifications: First, because the discrete Fourier
transform is linear, a transform of a mixture is simply a mixture of
the corresponding transforms. Second, the fact that the original trees
satisfy a set of phylogenetic invariants reduces the complexity of the
mixed invariants. Finally, the product of the exponentials of the
branch lengths appear in all formulas, and division leads to a
substantial simplification.

First we remind the reader of the main tools and fix notation.
Note that for the entire paper we will be working with the two-state
symmetric (also known as Cavender-Farris-Neyman) model.

\SBsubsection{The Hadamard transform and phylogenetic invariants}

\ For a given edge $e$ of branch length $\gamma(e)$ we will denote 
\begin{equation}
\theta(e) = \exp(-2 \gamma(e))
  \label{eq:fidelity_edgelength}
\end{equation}
which ranges between zero and one for positive branch lengths.
We call this number the ``fidelity'' of the edge, as it quantifies the
quality of transmission of the ancestral state across the edge.
For $A \subseteq \{1,\ldots,n\}$ of even order, let $q_A = \left( H_{n-1} \bar{p} \right)_A$ be the Fourier
transform of the split probabilities, where $H_n$ is the $n$ by $n$
Hadamard matrix \citep{semple-steel}.

Quartet trees will be designated by their splits, i.e. $13|24$
refers to a quartet with taxa labeled 1 and 3 on one side of the
quartet and taxa 2 and 4 on the other.  

By the first identity in the proof of Theorem~8.6.3 of 
\citep{semple-steel} one can express the Fourier
transform of the split probabilities in terms of products of
fidelities. That is, for any subset $A \subseteq \{1,\ldots,n\}$ of even order, 
\begin{equation}
q_A = \prod_{e \in \paths(T,A)} \theta(e)
\label{eq:path_prod}
\end{equation}
where $\paths(T,A)$ is the set of edges which lie in the set of
edge-disjoint paths
connecting the taxa in $A$ to each other. This set is uniquely
defined (again, see \citep{semple-steel}).

From this equation, we can derive values for the fidelities from the
Fourier transforms of the split probabilities. In particular, it is
simple to write out the fidelity of a pendant edge on a quartet. For
example, 
\[
\theta_1 = \sqrt{\frac{\theta_1 \theta_5 \theta_4 \cdot \theta_1
\theta_2}{\theta_2 \theta_5 \theta_4}} = \sqrt{\frac{q_{14} \ 
q_{12}}{q_{24}}}
\]
for a tree of topology $12|34$.
In general, we have the following lemma:
\begin{lem}
  If $a$, $b$, and $c$ are distinct pendant edge labels on a quartet
  such that $a$ and $b$ are adjacent, then the fidelity of a pendant
  edge $a$ is
  \[
\sqrt{\frac{q_{ab} \ q_{ac}}{q_{bc}}}.
  \]
  \label{lma:pendant_fidel}
\end{lem}
A similar calculation leads to an analogous lemma for the internal
edge:
\begin{lem}
  The fidelity of the internal edge of an $ab|cd$ quartet
  tree is
  \[
  \sqrt{\frac{q_{ac} \ q_{bd}}{q_{ab} \ q_{cd}}}.
  \]
  \label{lma:internal_fidel}
\end{lem}

This paper will also make extensive use of the method of phylogenetic
invariants. These are polynomial identities in the Fourier transform of the split
probabilities which are satisfied for a given tree topology.
Invariants are understood in a very general
setting (see \citet{pmid15767777}), however here we only require invariants
for the simplest case: a quartet tree with the two-state symmetric
model.  In particular, for the quartet tree $ab|cd$, the two phylogenetic
invariants are
\begin{eqnarray}
  & q_{abcd} - q_{ab} \ q_{cd} = 0 & \label{eq:invar1} \\
  & q_{ac} \ q_{bd} - q_{ad} \ q_{bc} = 0. & \label{eq:invar2}
\end{eqnarray}
A $q$-vector mimics the Fourier transforms of site pattern
frequencies of a nontrivial tree
exactly when they satisfy the phylogenetic invariants and have
corresponding edge fidelities (given by Lemmas \ref{lma:pendant_fidel}
and \ref{lma:internal_fidel}) between zero and one.

This paper is primarily concerned with the following situation: a mixture of two sets of
branch lengths on a quartet tree which mimics the site pattern frequencies of an
unmixed tree. We fix the following notation: the two branch length sets
will be called $t_i$ and $s_i$, the corresponding fidelities will
be called $\theta_i$ and $\psi_i$, and the Fourier transforms of the
site pattern frequencies will be labeled with $q$ and $r$,
respectively. The internal edge of the quartet will carry the label
$i=5$, and the pendant edges are labeled according to their terminal
taxa (e.g. $i=2$ is the edge terminating in the second taxon). The
mixing weight will be written $\alpha$, and we make the convention
that the mixture will take the $t_i$ branch length set with probability
$\alpha$ time and $s_i$ with probability $1-\alpha$. 

\SBsubsection{Mixtures mimicking a tree of the same topology}

\ In this section we describe conditions on mixtures such that a
nontrivial mixture of two branch lengths on $12|34$ can give the same probability
distribution as a single tree of the same topology. 

Mixing two branch length sets on a $12|34$ quartet tree with the above
notation leads to the following form of invariant
(\ref{eq:invar1}) for a resulting tree also of topology $12|34$:
\begin{equation}
\begin{split}
(\alpha + 1-\alpha) (\alpha \, q_{1234} + (1-\alpha) \, r_{1234}) - \\
(\alpha \, q_{12} + (1-\alpha) \, r_{12})
(\alpha \, q_{34} + & (1-\alpha) \, r_{34}) = 0.
\end{split}
\label{eq:sm_inv1a}
\end{equation}
Multiplying out terms then collecting, there will be a 
$\alpha^2 (q_{1234} - q_{12} q_{34})$ term which is zero
by the phylogenetic invariants for the $12|34$ topology. Similarly, the
terms with $(1-\alpha)^2$ vanish. Dividing by $\alpha \, (1-\alpha)$
which we assume to be nonzero, equation (\ref{eq:sm_inv1a}) becomes
\[
q_{1234} + r_{1234} - (q_{12} r_{34} + r_{12} q_{34}) = 0.
\]
Applying invariant (\ref{eq:invar1}) for the $12|34$ topology and simplifying leads to the
following equivalent form of (\ref{eq:sm_inv1a}):
\begin{equation}
(q_{12} - r_{12}) (q_{34} - r_{34}) = 0.
\label{eq:sm_inv1b}
\end{equation}
The same sorts of moves lead to the second invariant of the mixed
tree: 
\begin{equation}
q_{13} r_{24} + r_{13} q_{24} - (q_{14} r_{23} + r_{14} q_{23}) = 0.
\label{eq:sm_inv2b}
\end{equation}
The fact that $\alpha$ doesn't appear in these equations already
delivers an interesting fact: if a mixture of two branch lengths in this
setting satisfy the phylogenetic invariants for a single $\alpha$, then 
they do so for all $\alpha$. Geometrically, this means if the line
between two points on the subvariety cut out by the phylogenetic
invariants intersects the subvariety non trivially then
it sits entirely in the subvariety. 

We can gain more insight by considering these equations in terms of
fidelities. Direct substitution using (\ref{eq:path_prod}) into
(\ref{eq:sm_inv1b}) gives 
\[
(\theta_1 \theta_2 - \psi_1 \psi_2) (\theta_3 \theta_4 - \psi_3 \psi_4) = 0. 
\]
This equation will be satisfied exactly when the branch lengths satisfy 
\begin{equation}
t_1 + t_2 = s_1 + s_2 \ \ \hbox{or} \ \ t_3 + t_4 = s_3 + s_4.
\label{eq:sm_inv1c}
\end{equation}
The corresponding substitution into (\ref{eq:sm_inv2b}) 
and then division by $\theta_2 \theta_5 \theta_4 \psi_2 \psi_5 \psi_4$
gives after simplification
\[
\left( \frac{\theta_1}{\theta_2} - \frac{\psi_1}{\psi_2} \right)
\left( \frac{\theta_3}{\theta_4} - \frac{\psi_3}{\psi_4} \right) = 0
\]
This equation will be satisfied exactly when the branch lengths satisfy 
\begin{equation}
t_1 - t_2 = s_1 - s_2 \ \ \hbox{or} \ \ t_3 - t_4 = s_3 - s_4.
\label{eq:sm_inv2c}
\end{equation}

To summarize,
\begin{prop}
  \label{prop:same_mix}
  The mixture of two $12|34$ quartet trees with pendant branch lengths $t_i$ and
  $s_i$ satisfies the $12|34$ phylogenetic invariants for the binary
  symmetric model exactly (up to renumbering) when
  either $t_1 = s_1$ and $t_2 = s_2$, or 
$t_1 + t_2 = s_1 + s_2$ and $t_3 - t_4 = s_3 - s_4$. 
\end{prop}
As described above this proposition makes no reference to the mixing
weight $\alpha$.

In quartets where $t_1 = s_1$ and $t_2 = s_2$, the resulting tree will
also have pendant branch lengths $t_1$ and $t_2$:
\begin{prop}
  \label{prop:mix_preserves}
  A mixture of two $12|34$ quartet trees with branch lengths $t_i$ and
  $s_i$ which satisfies $t_1 = s_1$ and $t_2 = s_2$ will have
  resulting pendant branch lengths for the first and second taxa equal to
  $t_1$ and $t_2$, respectively.
\end{prop}

\begin{proof}

  Let the fidelity of the edges leading to taxon one and two be
  denoted $\mu_1$ and $\mu_2$. We have by Lemma~\ref{lma:pendant_fidel} with
  $a=1$, $b=2$ and $c=3$,
  \[
 \mu_1 =
\sqrt{\frac{
(\alpha \theta_1 \theta_2 + (1-\alpha) \psi_1 \psi_2) \cdot 
(\alpha \theta_1 \theta_5 \theta_3+(1-\alpha) \psi_1
\psi_5 \psi_3) } {\alpha \theta_2 \theta_5 \theta_3 + (1-\alpha) \psi_2 \psi_5
\psi_3}}
  \]
This fraction is equal to $\theta_1$ after substituting $\psi_1 =
\theta_1$ and $\psi_2 = \theta_2$, which are implied by the
hypothesis. The same calculation implies that $\mu_2 = \theta_2$. 
\end{proof}

In the rest of this section we note that anomalous branch lengths can
emerge from mixtures of trees mimicking a tree of the same topology.
\begin{prop}
It is possible to mix two sets of branch lengths on a
tree to mimic a tree of the same topology such that one resulting pendant
branch length is arbitrarily small while the corresponding branch length in
either of the branch length sets being mixed stays above some arbitrarily
large fixed value.
\end{prop}
 
\begin{proof}
To get such an anomalous mixture, set $\theta_1 = \psi_1$, $\theta_3 =
\psi_3$, $\theta_4 = \psi_4$, $\theta_2 = \psi_5$, $\theta_5 =
\psi_2$, and $\alpha = .5$. The equations (\ref{eq:sm_inv1c}) and
(\ref{eq:sm_inv2c}) are satisfied because $\theta_3 = \psi_3$ and $\theta_4
= \psi_4$, and therefore $t_3 = s_3$ and $t_4 = s_4$. This implies
that the mixture will indeed satisfy the phylogenetic invariants.

Now, because again the Fourier transform of a mixture
is the mixture of the Fourier transform, using
Lemma~\ref{lma:pendant_fidel} and simplifying gives
\begin{equation}
\mu_1 = \frac{ \theta_1 | \theta_2 + \theta_5 |}{\sqrt{\theta_2
\theta_5}}
\label{eq:mu_mix}
\end{equation}

Now note that by making the ratio $\theta_2 / \theta_5$ small,
it is possible to have $\mu_1$ be close to one although
$\theta_1$ can be small. This setting corresponds (via
(\ref{eq:fidelity_edgelength})) to the case of the first branch length of
the resulting tree to be going to zero although the trees used to
make the mixture may have long first branch lengths. It can be checked 
by calculations analogous to (\ref{eq:mu_mix}) that the other
fidelities of the tree resulting from mixing will be, in order,
$\sqrt{\theta_2 \theta_5}$, $\theta_3$, $\theta_4$, $\sqrt{\theta_2
\theta_5}$. These are clearly strictly between zero and one, so the
resulting tree will have positive branch lengths.
\end{proof}

\SBsubsection{Mixtures mimicking a tree of a different topology}

\ In this section we answer the question of what branch lengths on a quartet
can mix to mimic a quartet of a different topology. 

\begin{prop}
  \label{prop:main}
Let $k_1,\ldots, k_4$ satisfy the following inequalities:
\begin{eqnarray}
  & k_1 > k_3 > k_4 > 1 > k_2 > 0, \label{eq:IVd} & \\
  & \frac{1-k_1^2}{k_1} \frac{1-k_4^2}{k_4} + \frac{1-k_2^2}{k_2}
  \frac{1-k_3^2}{k_3} > 0, \label{eq:IVa} & \\
  & \frac{k_1+k_4}{1+k_1 k_4} \cdot \frac{k_2+k_3}{1+k_2 k_3} > 1.
  \label{eq:IVc} &
\end{eqnarray}
Then there exists $\pi_5$ such that for any $\pi_5 < k_5 < \pi_5^{-1}$
sufficiently close to either $\pi_5$ or $\pi_5^{-1}$ there exists a
mixing weight such that for any $t_1,\ldots,t_5$ and $s_1,\ldots,s_5$
satisfying $\pi_5 = \exp\left(-2(t_5 + s_5)\right)$ and $k_i = \exp\left(-2(t_i - s_i)\right)$ for $i=1,\dots,5$, 
the
corresponding mixture of two $12|34$ trees will satisfy the
phylogenetic invariants for a single tree of the $13|24$ topology. The
resulting internal branch length is guaranteed to be positive, and the
pendant branch lengths will be positive as long as the pendant
branch lengths being mixed are sufficiently large.
\end{prop}

\begin{proof}
Let $m$ denote the Fourier transform vector of the site pattern
frequencies of the mixture.
The invariants for a tree of topology $13|24$ are (by
(\ref{eq:invar1}) and (\ref{eq:invar2}))
\begin{eqnarray}
  & m_{1234} - m_{13} m_{24} = 0 & \label{eq:13invar1} \\
  & m_{12} m_{34} - m_{14} m_{23} = 0 & \label{eq:13invar2} .
\end{eqnarray}

As before, we insert the mixture of the Fourier transforms of the
pattern frequencies into the invariants. For the first invariant, 
\[
\begin{split}
(\alpha + 1-\alpha) (\alpha \, q_{1234} + (1-\alpha) \, r_{1234}) \\
- (\alpha \, q_{13} + (1-\alpha) \, r_{13})
(\alpha \, q_{24} + & (1-\alpha) \, r_{24}) = 0.
\end{split}
\]
Multiplying, this is equivalent to 
\begin{equation}
\label{eq:di_inv1a}
\begin{split}
  \alpha^2 (q_{1234} - q_{13} q_{24}) \\
  + \alpha (1-\alpha) & \left(q_{1234} + r_{1234} - (q_{13} r_{24} + r_{13} q_{24}) \right) \\ 
  + (1-\alpha)^2 & (r_{1234} - r_{13} r_{24}) = 0.
\end{split}
\end{equation}

A similar calculation with the second invariant leads to 
\begin{equation}
\label{eq:di_inv2a}
  \begin{split}
  \alpha^2 (q_{12} q_{34} - q_{14} q_{23}) \\ 
  + \alpha (1 - \alpha) & \left(q_{12} r_{34} + r_{12} q_{34} - (q_{14} r_{23} + r_{14} q_{23}) \right) \\ 
  + (1-\alpha)^2 & (r_{12} r_{34} - r_{14} r_{23}) = 0.
  \end{split}
\end{equation}

Rather than (\ref{eq:di_inv1a}) and (\ref{eq:di_inv2a}) themselves, we
can take (\ref{eq:di_inv1a}) and the difference of (\ref{eq:di_inv1a})
and (\ref{eq:di_inv2a}). 
Because the $q$ and $r$ vectors come from a tree with topology
$12|34$, they satisfy 
$ q_{1234} = q_{12} q_{34}$ and $q_{13} q_{24} = q_{14} q_{23}$ and
the equivalent equations for the $r$.
Thus the difference of (\ref{eq:di_inv1a}) and (\ref{eq:di_inv2a}) can
be simplified to (assuming $\alpha (1-\alpha) \neq 0$)
\begin{equation}
  \begin{split}
  q_{1234} + r_{1234} - (q_{12} r_{34} + r_{12} q_{34}) \\
  = q_{13} r_{24} + r_{13} q_{24} - & (q_{14} r_{23} + r_{14} q_{23}).
  \end{split}
  \label{eq:di_inv2b}
\end{equation}

We would like to ensure that the tree coming from the mixture has
nonzero internal branch length. By Lemma~\ref{lma:internal_fidel} this is
equivalent to showing that 
\begin{equation}
  m_{13} \ m_{24} > m_{14} \ m_{23}.
  \label{eq:nontriv_internal}
\end{equation}
Substituting in for
the mixture fidelities and simplifying results in 
\[
\begin{split}
\alpha^2 (q_{13} q_{24} - q_{14} q_{23}) \\
+ \alpha (1-\alpha) & \left(q_{13} r_{24} + r_{13} q_{24} - (q_{14} r_{23} + r_{14} q_{23}) \right) \\ 
+ (1-\alpha)^2 & (r_{13} r_{24} - r_{14} q_{23}) > 0.
\end{split}
\]
The first and last terms of this expression vanish because the
$q$ and $r$ satisfy the $12|34$ phylogenetic
invariants coming from (\ref{eq:invar1}) and (\ref{eq:invar2}). Simplifying leads
to 
\begin{equation}
  q_{13} r_{24} + r_{13} q_{24} > q_{14} r_{23} + r_{14} q_{23}.
  \label{eq:di_ela}
\end{equation}

Define $k_i = \psi_i / \theta_i$ for $i=1,\dots,5$ and $\rho = \alpha
/ (1 - \alpha)$. Note that 
\begin{equation}
  0 < \theta_i < \min(k_i^{-1},1)
  \ \hbox{and} \ 
  0 < k_i < \infty 
  \label{eq:k_i_restr}
\end{equation}
 is equivalent to $0 < \theta_i < 1$ and
$0 < \psi_i < 1$.  Define 
\begin{eqnarray*}
  \chi_{12} = k_1 k_2 + k_3 k_4  & & \chi_{13} = k_1 k_3 + k_2 k_4 \\
  \chi_{14} = k_1 k_4 + k_2 k_3 & & \chi_{1234} = 1+ k_1 k_2 k_3 k_4.
\end{eqnarray*}
Later we will make use of the fact that the $\chi$ are
invariant under the action of the Klein four group. 

Using these definitions, direct substitution using (\ref{eq:path_prod}) into (\ref{eq:di_inv1a}),
(\ref{eq:di_inv2b}), and (\ref{eq:di_ela}) and some simplification
shows that the set of equations
\begin{eqnarray}
  & \begin{split}
    \rho^2 (1-\theta_5^2) + \rho(\chi_{1234} - \theta_5 \psi_5
    \chi_{13}) \\
    + (1-\psi_5^2) & (\chi_{1234} - 1) = 0 \end{split} &
  \label{eq:IIa} \\
  & \chi_{1234} - \chi_{12} = \theta_5 \psi_5 (\chi_{13} - \chi_{14}) &
  \label{eq:IIb} \\
  & \chi_{13} > \chi_{14} &
  \label{eq:IIc}
\end{eqnarray}
is equivalent to equations (\ref{eq:13invar1}), (\ref{eq:13invar2})
and (\ref{eq:nontriv_internal}).

Equation (\ref{eq:IIb}) is simply satisfied by setting
\begin{equation}
  \theta_5 \psi_5 = 
  \frac{\chi_{1234} - \chi_{12}}{\chi_{13} - \chi_{14}}.
  \label{eq:t5p5}
\end{equation}
However, in doing so, we must require that this ratio is strictly
between zero and one. The fact that it must be less than one can be written
\begin{equation}
  \chi_{14} + \chi_{1234} < \chi_{12} + \chi_{13}
  \label{eq:IIIc}
\end{equation}
which by a short calculation is equivalent to (\ref{eq:IVc}).
Later it will be shown that other equations imply that
(\ref{eq:t5p5}) is greater than zero.

Assign variables $A$, $B$, and $C$ in the standard way such that 
(\ref{eq:IIa}) can be written $A \rho^2 + B \rho + C$. 
The $A$ and $C$ terms are strictly positive, thus
the existence of a $0 < \rho < \infty$ satisfying this equation
implies 
\begin{equation}
  B < 0 \ \hbox{and} \ B^2 - 4 A C > 0. 
  \label{eq:quad}
\end{equation}
On the other hand, (\ref{eq:quad}) implies the existence of a 
$0 < \rho < \infty$ satisfying (\ref{eq:IIa}).

Note that using (\ref{eq:t5p5}), $B<0$ is equivalent to
\[
\chi_{1234} - \frac{\chi_{1234} - \chi_{12}}{\chi_{13} - \chi_{14}}
\chi_{13} < 0.
\]
Multiplying by $\chi_{13} - \chi_{14}$ which is positive by
(\ref{eq:IIc}) this equation is equivalent to 
\begin{equation}
  \chi_{12} \chi_{13} < \chi_{1234} \chi_{14}
  \label{eq:IIIa}
\end{equation}
which by a short calculation is equivalent to (\ref{eq:IVa}).
The conclusion then is that the existence of a $\rho \geq 0$
satisfying (\ref{eq:IIa}) is equivalent to
(\ref{eq:IVa}) and $B^2 - 4AC > 0$ given the rest of the invariants. 

Now, (\ref{eq:IIc}) and (\ref{eq:IIIa}) imply that $\chi_{12} <
\chi_{1234}$. Therefore, according to (\ref{eq:t5p5}) the product
$\theta_5 \psi_5$ is greater than zero given (\ref{eq:IIc}). 
For convenience, set $\pi_5 = \theta_5 \psi_5$, which as described is
determined by $k_1, \dots, k_4$. Now, $\theta_5$ being less than
one and $\psi_5$ being less than one are equivalent to
\begin{equation}
  \pi_5 < k_5 < \pi_5^{-1}.
  \label{eq:k5_restr}
\end{equation}
In summary, the problem of finding branch lengths and a mixing parameter such that the derived
variables satisfy (\ref{eq:13invar1}), (\ref{eq:13invar2}) and
(\ref{eq:nontriv_internal})
is equivalent to finding $k_i$ and
$\theta_i$ satisfying (\ref{eq:IVa}), (\ref{eq:IVc}), (\ref{eq:k_i_restr}), (\ref{eq:IIc}),
(\ref{eq:t5p5}), (\ref{eq:k5_restr})
  and $B^2 - 4AC > 0$, which can be written
\begin{equation}
  (\chi_{1234} - \pi_5 \chi_{13})^2 - 4 (1 - \pi_5 / k_5) (1-\pi_5
  k_5) (\chi_{1234}-1) > 0. 
  \label{eq:IIIb}
\end{equation}
Note that $\chi_{1234} = \pi_5 \chi_{13}$ is impossible using
(\ref{eq:IIb}) and (\ref{eq:IIIa}).
Therefore (\ref{eq:IIIb}) can be satisfied while fixing the other
variables by taking $k_5$ close to $\pi_5$ or $\pi_5^{-1}$ while
satisfying (\ref{eq:k5_restr}).

Now we show that (possibly after relabeling) equation (\ref{eq:IVd}) is equivalent to
(\ref{eq:IIc}) in the presence of the other inequalities.
Recall that the $\chi$ are invariant under the action of
the Klein group acting on the indices of
$k_i$.  Because the invariants are equivalent to equations which can
be expressed in terms of the $\chi$ with $\theta_5$ and
$\psi_5$, we can assume that $k_1 \geq k_2$ and $k_1 \geq
k_3$ by renumbering via an element of the Klein group. 

Now, subtract $\chi_{12} \chi_{14}$ from (\ref{eq:IIIa}) to find 
\[
\chi_{12} ( \chi_{13} - \chi_{14} ) < (\chi_{1234} - \chi_{12} )
\chi_{14}.
\]
Rearranging (\ref{eq:IIIc}), it is clear that this implies that
\begin{equation}
  \chi_{12} < \chi_{14}.
  \label{eq:chi12_14}
\end{equation}
Inserting the definition of the $\chi$ into (\ref{eq:IIc}) and
(\ref{eq:chi12_14}) shows that these equations are equivalent to
\begin{equation}
  0 < (k_1 - k_2) (k_3 - k_4) \ \hbox{and} \ 0 < (k_1 - k_3) (k_4 - k_2). 
  \label{eq:k_rel_ranks}
\end{equation}
We have assumed by symmetry that $k_1 \geq k_2$ and $k_1 \geq k_3$;
now (\ref{eq:k_rel_ranks}) shows that $k_1$ can't be equal to either
$k_2$ or $k_3$.  Also, (\ref{eq:k_rel_ranks}) shows that $k_3 > k_4$ and
$k_4 > k_2$.  All of these inequalities put together imply that $k_1 >
k_3 > k_4 > k_2$, which directly implies (\ref{eq:IIc}).

Furthermore, another rearrangement of (\ref{eq:IIIc}) using the
inequality (\ref{eq:chi12_14}) leads to $\chi_{1234} < \chi_{13}$.
This after substitution gives $(1-k_1 k_3) (1-k_2 k_4) < 0$, which
implies that it is impossible for all of the $k_i$ to be either less than
or greater than one. 

Note that (\ref{eq:IVa}) 
excludes the case $k_1 > k_3 > 1 > k_4 > k_2$; this leaves $k_1 > 1 >
k_3 > k_4 > k_2$ and $k_1 > k_3 > k_4 > 1 > k_2$. We can assume the
latter without loss of generality by exchanging the $\theta_i$ and the
$\psi_i$ (which corresponds to replacing $k_i$ with $k_i^{-1}$) and
renumbering.

So far we have described how to find values for the branch lengths so
that the invariants (\ref{eq:invar1}) and (\ref{eq:invar2}) and the
internal branch length inequality (\ref{eq:nontriv_internal}) are
satisfied. However, we also need to check that the resulting
pendant branch lengths for the tree are positive. Here we describe how this can
be achieved by taking a lower bound on the values of $t_i$. 

Assume edges $a$ and $b$ are adjacent on the $12|34$ trees being
mixed, and $a$ and $c$ are adjacent on the resulting $13|24$ tree.
Then, by Lemma~\ref{lma:pendant_fidel} and (\ref{eq:path_prod}), the
fidelity of the pendant $a$ edge is
\[
\sqrt{\frac{(\alpha \theta_a \theta_b + (1-\alpha) \psi_a \psi_b)
(\alpha \theta_a \theta_5 \theta_c + (1-\alpha) \psi_a \psi_5
\psi_c)}{\alpha \theta_b \theta_5 \theta_c + (1-\alpha) \psi_b \psi_5
\psi_c}}. 
\]
In order to assure that the resulting pendant branch length for edge $a$
is positive, we must show that the above fidelity is less
than one. This is equivalent to showing that $\theta_a$ must
satisfy
\begin{equation}
  \label{eq:theta_ub}
  \theta_a < 
  \sqrt{\frac{\alpha + (1-\alpha) k_b k_5 k_c}{(\alpha+(1-\alpha) k_a k_b) (\alpha + (1-\alpha) k_a
  k_5 k_c)}}
\end{equation}
for all such $a$, $b$, $c$ triples.
Thus this equation along with (\ref{eq:k_i_restr}) imply
upper bounds for $\theta_a$; by the definition of fidelities these
translate to lower bounds for $t_a$.  This concludes the proof.
\end{proof}

Note that the proof actually completely characterizes (up to
relabeling) the set of branch lengths and mixing weights such that the
resulting mixture mimics a
tree of different topology.  

\begin{prop}
  If two sets of branch lengths on the $12|34$ tree mix to mimic a tree of
  the topology $13|24$ then up to relabeling the associated $k_i$ must
  satisfy the inequalities (\ref{eq:IVd}), (\ref{eq:IVa}),
  (\ref{eq:IVc}), and (\ref{eq:k5_restr}); the $\theta_i$ must satisfy
  the inequalities (\ref{eq:k_i_restr}) and (\ref{eq:theta_ub}). The
  two required equalities are that the product $\theta_5 \psi_5$ must
  satisfy (\ref{eq:t5p5}), and the associated $\rho$ must satisfy
  (\ref{eq:IIa}).  
\end{prop}

\SBsubsubsection{Kullback-Leibler lemma}
\begin{lem}
  Assume some group-based model $G$ and let $\ps$ be the
  probability simplex for distributions on four taxa under
  $G$. Let $V \subset \ps$ be the set of all site-pattern frequencies for some
  quartet tree under $G$. Then 
  \[
  \dkl(p, V) := \min_{v \in V} \dkl (p,v)
  \]
  exists and is continuous for all $p$ in the interior of $\ps$.
\label{lem:kl_cont}
\end{lem}

\begin{proof}

Note that $\dkl(p,q)$ is a continuous function when probability
distributions $p$ and $q$ have no components zero, i.e. they sit in the
interior $\psint$ of the probability simplex $\ps$.
We will show that for any $p \in \psint$ there exists an open
neighborhood $U$ of $p$ such that $\dkl(p',V)$ exists and is continuous for
all $p' \in U$. 
Given $p$ let $\pmin$ be the smallest component $p_i$ of $p$. Let
$U = \left\{ p' \in \psint : p_i' > \pmin / 2 \right\}.$
Then choose $\varepsilon > 0$ such that 
\[
\log(\pmin / 2) + \frac{1}{2} \, \pmin \, \log ( 1/ \varepsilon ) > 
\sup_{p' \in U} \inf_{q \in V} \dkl(p',q).
\]
The right hand side of this equation is finite (since it is bounded
above by $\sup_{p' \in U} \delta_{\rm{KL}}(p', q^*)$ for any point $q^*
\in V$ with no components zero).

Let $B = \{q \in V : \, q_i \geq \varepsilon \  \hbox{ for all } i \}$.
$V$ is a compact set \citep{MR2095862} therefore $B
\subset \psint$ is compact as well. 
Now for any $p' \in U$ and $q' \in V-B$
\begin{eqnarray*}
  \dkl(p', q') & = & \sum_{i} p_i' \log p_i' + \sum_{i} p_i' \log (1/q_i') \\
  & > & \log(\pmin / 2) + \frac{1}{2} \pmin \log ( 1/ \varepsilon ) \\
  & > & \inf_{q \in V} \dkl(p',q)
\end{eqnarray*}
so the infimum cannot be achieved outside $B$. Consequently,
\[
\inf_{q \in V} \dkl (p',q) = \min_{q \in B} \dkl(p',q)
\]
for all $p' \in U$. 
Thus the right hand side exists; continuity follows from standard
analytic arguments.
\end{proof}


\newpage 
\noindent
Table 1: Rounded branch lengths for the examples in Figure~1.  The top
division of the table is example (a); the bottom is example (b). The
top two lines in each division are the branch lengths forming the mixture
and the third line gives the branch lengths for the unmixed tree.

\vspace{1cm}

\noindent
\begin{table}[h]
  \begin{center}
    \begin{tabular}{l|lllll}
      weight & pendant 1 & pendant 2 & pendant 3 & pendant 4 & internal \\ 
      \hline
      \hline
      0.748646 & 1.772261 & 0.25 & 0.949306 & 0.846574 & 0.366516 \\
      0.251354 & 0.25 & 1.353637 & 0.4 & 0.5 & 0.213387 \\
      1. & 0.888101 & 0.905792 & 0.648625 & 0.654236 & 0.086051\\
      \hline
      \hline
      0.936064 & 1.838398 & 0.2 & 1.397309 & 0.411489 & 0.062429 \\
      0.063936 & 0.2 & 0.543932 & 0.2 & 0.2 & 0.055312 \\
      1. & 1.011471 & 0.375718 & 0.794529 & 0.305338 & 0.360827\\
      \hline
    \end{tabular}
  \end{center}
\end{table}


\newpage 
\noindent
Figure 1: Mixtures of two sets of branch lengths on a tree of a given
topology can have exactly the same site pattern frequencies as a tree
of a different topology under the two-state symmetric model.  The
notation in the diagram showing $x * T_1 + (1-x) * T_1^{'} = T_2$
means that the indicated mixture of the two branch lengths sets $T_1$
and $T_1'$ shown
in the diagram gives the same expected site pattern frequencies as the
tree $T_2$.  The diagrams show two examples of this ``mixed branch
repulsion;'' the general criteria for such mixtures is explained in
the text.  The branch length scale in the diagrams is given by the
line segment indicating the length of a branch with 0.5 substitutions
per site. Note that the mixing weights in this example have been
rounded.
\vspace{1cm}
\newpage 

\noindent
Figure 2: Mixtures of two sets of branch lengths on a tree of a given
topology can have exactly the same site pattern frequencies as a tree
of the same topology under the two-state symmetric model. The
criterion for the occurrence of this phenomenon is explained in the
text and an example is shown in the figure. Note in particular that
the branch lengths need not average: for example, the branch length
for the pendant edge leading to taxon $1$ virtually disappears after
mixing.
\vspace{1cm}
\newpage 

\noindent
Figure 3: A geometric depiction of the main result. The ambient space
is a projection of the seven-dimensional probability simplex of site
pattern frequencies for trees on four leaves. The gray sheet is a
subset of a two-dimensional subvariety of the site pattern
frequencies for trees of the $12|34$ topology, while the black sheet
is an analogous subset for the $13|24$ topology. The horizontal line
represents the possible mixtures for the two sets of branch lengths
for the $12|34$ topology in Figure 1a. The fact that these two sets of
branch lengths can mix to make a tree of topology $13|24$ is shown
here by the fact that the horizontal line intersects the black sheet.
\vspace{1cm}

\end{spacing}

\end{sysbiol}
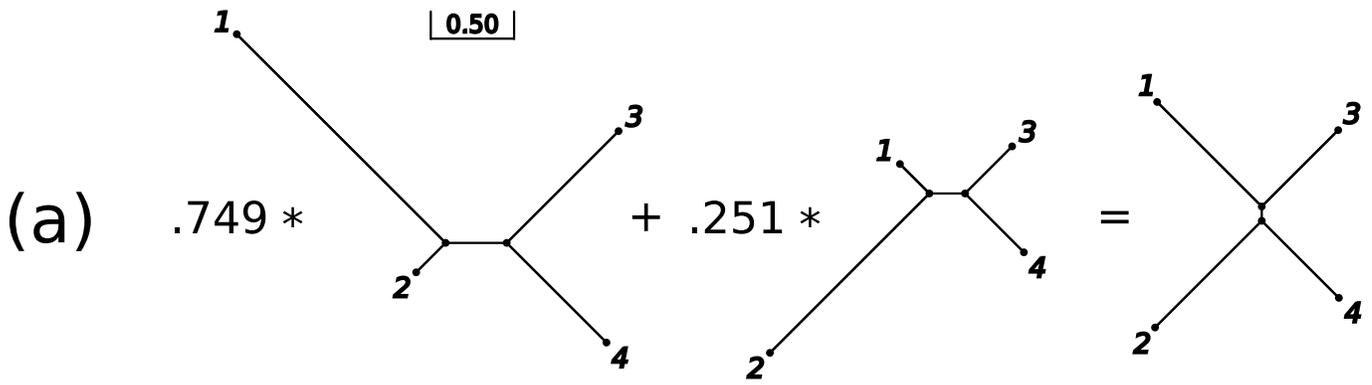
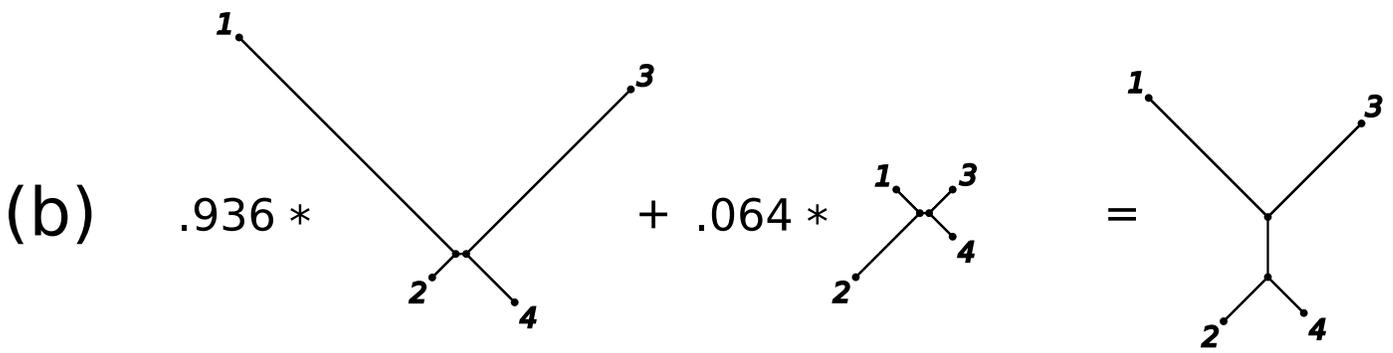
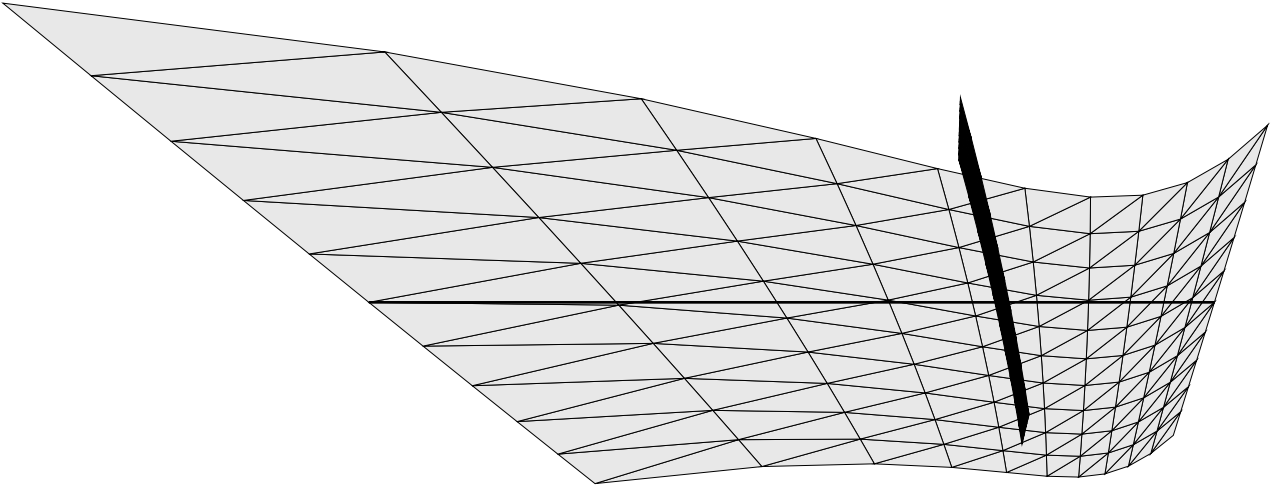
\end{document}